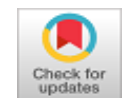

# Estimation of Similarity between DNA Sequences and Its Graphical Representation

**Probir Mondal**

*Abstract*: *Bioinformatics, which is now a well known field of study, originated in the context of biological sequence analysis. Recently graphical representation takes place for the research on DNA sequence. Research in biological sequence is mainly based on the function and its structure. Bioinformatics finds wide range of applications specifically in the domain of molecular biology which focuses on the analysis of molecules viz. DNA, RNA, Protein etc. In this review, we mainly deal with the similarity analysis between sequences and graphical representation of DNA sequence.*

*Keywords:* DNA sequence, Gene expression, *Graphical representation*, *Similarity analysis,*

## I. INTRODUCTION

Information Science and technology is take part into modern molecular biology since the last few decades. Developing tools and applications are the recent focus on biological sequence analysis. When our area of study is biological sequence then the word 'bioinformatics' has to come. Bioinformatics [1, 2] means the most important digital biological information which arises after analysis on biological sequences. Applications of Computer science and information technology is take place in the field of molecular biology recently. Bioinformatics is being used in various fields like Forensic analysis, Evolutionary studies, Molecular medicine and Biotechnology etc. Biotechnology combines disciplines like cell biology, biochemistry, genetics and molecular biology. It focuses on the analysis of the three molecules, viz. DNA, RNA and Protein. Among the three molecules we shall focus attention mainly on DNA.

DNA contains required information to construct and maintain an organism [3]. Each cell in an organism has a DNA and it instructs cell what proteins to make. The "double helix" structure of DNA was discovered in 1953 by Francis Crick and James Watson. Backbone of this structure is formed by sugar and phosphate. (See figure 1), that are connected by chemical bonds. One strand is in the direction of 5' - to 3' and other in the 3' to 5' direction. There are four nucleotides: Adenine, Thymine, guanine and cytosine. The ordered sequence [6, 7], of these bases determines the information available for building and maintaining an organism. There are two groups, Purines and Pyrimidines. Purines are double ringed structure with nine membered and pyrimidines are single ringed structure with six members. In purines, there are two nucleobases Adenine and Guanine and for pyrimidines, there are three nucleobases, Thymine (T), Cytosine (C), and Uracil (U) (See Figure 2). In DNA, Adenine bonds with Thymine and Guanine bonds with Cytosine, these bonds are hydrogen bond (see Figure 3).

RNA is another essential molecule; it is very similar to DNA. Structure of RNA is three dimensional and it consists of single-stranded, which contains sugar ribose and phosphate. In RNA, Thymine (T) is replaced by Uracil (U), it is another form of thymine. In Cellular organisms, RNA molecule take part a huge role to convey genetic information. Information is conveyed by messenger RNA (mRNA) to synthesis of specific proteins. Combination of four nucleotides (A, U, G and C) in three base pair represents 64 codons. Due to redundancy in genetic code, 64 codons are reduced to 20 amino acids, except the three stop codons. This process is known as transcription (see the figure 4). Before the mRNA can be used as a template for the productions of proteins it needs to be processed, this involves removing and adding section of RNA. The mRNA that moves out of the nucleus into the cytoplasm and it travel to a ribosome to binds the mRNA. Ribosome reads the code in the mRNA to produce a chain made up upon the amino acids. Transfer RNA molecule is carrying the amino acids to the ribosome. The mRNA read three bases at a time, as each triplet is read a transfer RNA delivers the corresponding amino acids. This is added to a growing chain of amino acids, this is known as polypeptide chains. Once the last amino acid has been added, the chain falls into a complex 3D shape to form the protein.

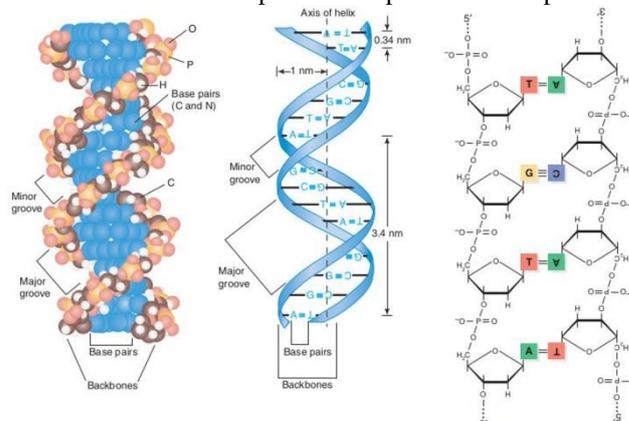

**Fig. 1.Structure of DNA with different molecules**

# Estimation of Similarity between DNA Sequences and Its Graphical Representation

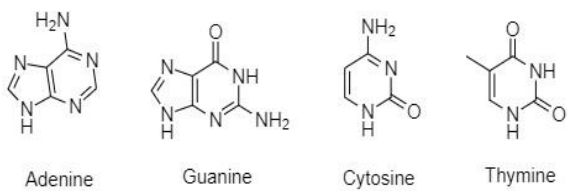

**Fig. 2. Structure of different nucleotides**

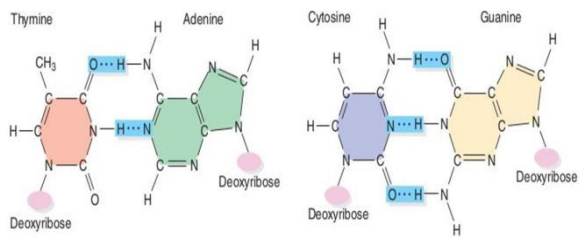

**Fig. 3. Bonding structure between Purine and Pyrimidine group**

**Fig. 4. Amino acid translation chart**

Use of graph theory in bioinformatics has been increased exponentially and day after day the strength of database of the biological sequences also increased. Different approaches are come to represent graph using biological sequences and determine similarity between two species. Our goal to accelerate the existing approaches by new proposals. Graphs are represents from biological sequences and obtaining statistical information to performing similarity analysis on it and it is initial domain of study. In our proposal, without loss of any biological information during the representation of graph [8] and analysis we should consider. Bioinformatics deals with variety of subjects but we primarily analyses the DNA and protein sequence and structures. Analysis of sequences may functional, structural or evolutionary.

## II. SEQUENCE ALIGNMENT

The first elemental tasks is finding parallelism between a pair of sequences. Several algorithms and tools are available in this domain for searching and alignment. We analyse sequences to find common patterns or similarity between them. If we discover such sequence similarities, we may infer biological similarity between two sequences. The relationships could be evolutionary or in terms of functional. Sometimes it could be structural also.

### A. Dot matrices

The first sequence comparison method is "dot-matrix analysis" [9, 10] or simply dot-plot and it published by Gibbs and McIntyre in 1970. It is a useful and simple tools for sequence comparison. The two sequences are written in such a way as to form a matrix. One sequence is written on top, from left to right, each small part in the sequence labeling one column of the matrix. the second sequence is written on the left, from top to bottom, each small part in the sequence labelling one row of the matrix. If two corresponding nucleotide base are identical then a dot is placed for each element of a matrix, else the element is left blank. After that we find diagonal rows of dots which indicates that two sequences are identical in that portion.

### B. Dynamic programming

Dynamic programming is a familiar technique to find the set of parameters that would give the desired optimum solution. There are several DP algorithms that have been developed for aligning two sequences. Here we deal with two of them - one global algorithm, namely the Needleman-Wunsch method and one local alignment method, namely the Smith Waterman method. In 1970, first application of dynamic programming to alignment of sequences published by Needleman and Wunsch in an elegant paper in the Journal of Molecular Biology[11]. In 1981, Smith and Waterman proposed a local alignment method [12], in which portions of one sequence are aligned to similar portions of the second sequence. It is useful in comparing a relatively short sequence with a very long sequence, or a set of long sequences. It frequently happens that a portion or portions of the relatively short query sequence is similar to not just one, but many regions of the long target sequence. Similarity of some portions of query sequence is aligned with target sequence with different order and different distance. The Smith-Waterman algorithm is designed to handle such cases.

### C. BLAST and FASTA

BLAST, FASTA, and other similarity searching analysis techniques are used to find homology between protein and DNA sequences. From few decades BLAST and FASTA provides very accurate and reliable result. Basic Local Alignment Search Tool (BLAST) [13-15] is the tool most frequently used for calculating sequence similarity compare to FASTA.

The modified version of the FASTP program is FASTA [16] program, which can be used as a sequence alignment tool like BLAST. Finding the maximum portion of related sequences in multiple portion of similarity is the advantage of FASTA.

## III. MULTIPLE ALIGNMENT

Secondly, we analyze the amount of similarity and dissimilarity on multiple sequences. Identifying parallelism between sequences is an important aspect on the ground of functional observations.

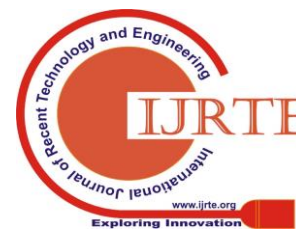

It can analyze further, for
phylogenetic analysis to identify relationships between an ancestral sequence and it descendants. SAT´e-II is a well known method for calculating phylogenetic alignments [17]. Graphical representation [4, 5] technique is convenient in case of sequence analysis and it used extensively used in the past. A graphical representation to describe DNA sequence was firstly introduced by Hamori and Ruskin [7] in the year 1983. Later there are several works are done in the past.

Some research works on multiple sequence alignment are explained.

**In 2010, Jia-Feng Yu** [18] et al. Sun proposed an approach to represent a given sequence into a graph, without loss of any biological information. To represent the graph they build up 16 kinds of dinucleotides from the four kind of bases A, G, C and T. Then distributed the dinucleotides into four quadrants in 2D (see figure 5 below).

**Table- I: Values of parameters for constructing the D–curve of the given sequence**

| Dinucleotides | a | b | k | c | a' | b' | c' |
|---|---|---|---|---|---|---|---|
| AT | 2 | 1 | 1 | 2 | 2 | 1 | 2 |
| TG | 1 | -2 | 2 | -2 | 3 | -1 | 0 |
| GG | -1 | 2 | 3 | -2 | 2 | 1 | -2 |
| GT | -2 | 1 | 4 | -2 | 0 | 2 | -4 |
| TG | 1 | -2 | 5 | -2 | 1 | 0 | -6 |
| GC | -2 | 2 | 6 | 4 | -1 | 2 | -10 |
| CA | -1 | -1 | 7 | 1 | -2 | 1 | -9 |
| AC | 2 | 2 | 8 | 4 | 0 | 3 | -5 |
| CC | -2 | -2 | 9 | 4 | -2 | 1 | -1 |

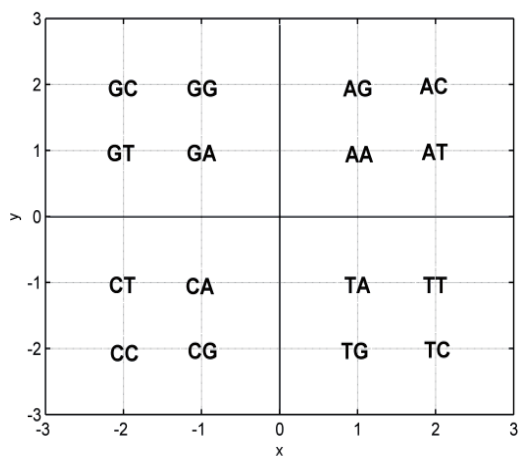

**Fig. 5.16 kinds of dinucleotides into four quadrants in 2D**

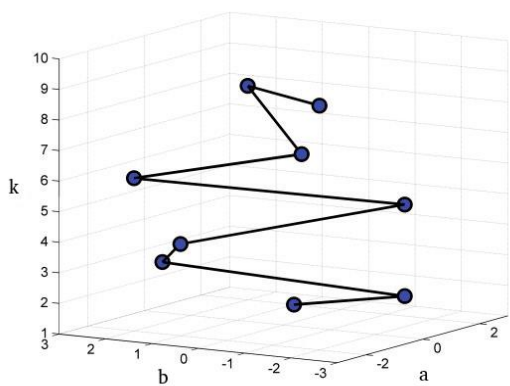

**Fig. 6. D – curve**

With sequence ATGGTGCACC, they easily find the value of a, b and k from the figure 5 and construct the table below. To calculate c they define c = a∗b and

$$a'_k = \sum_{i=1}^{k} a_i, \quad b'_k = \sum_{i=1}^{k} b_i, \quad c'_k = \sum_{i=1}^{k} c_i$$
$$k = 1,2,3,\dots,k-1 \quad (1)$$

Using the values of a, b and k we construct a D-curve (see figure 6). Similarity analysis is perform on 11 species based on the D-curve (consider only the β globin genes). Projection of the D-curve on 2-D a-b coordinates shown in Figure 7 and the figure 8 shown 2D curves based on $a_0$, $b_0$ and $c_0$. A closer observation in the Figure shows that Chimpanzee and gorilla have most aligned species and figure shows Human-Gorilla, Gorilla-Chimpanzee and Human-Chimpanzee are the most similar compositions. They also use Euclidian distance for analysis of similarity and dissimilarity. They find correlation between D-curve for further analysis of similarity. Use Pearson Correlated coefficient (PCC) for the same.

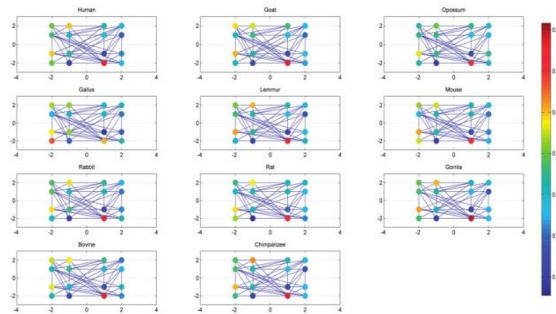

**Fig. 7. Projected D-curve on 2-D a-b coordinates**

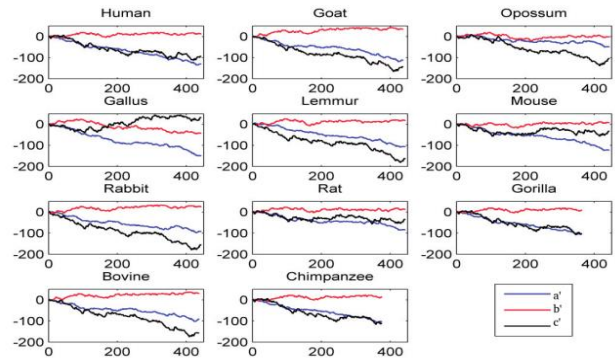

**Fig. 8. 2D curve based on a', b'and c'**

# Estimation of Similarity between DNA Sequences and Its Graphical Representation

**Zhujin Zhang** et al. introduced another new technique for graphical representation [19] of DNA sequences and also proposed a new approach for similarity analysis in 2010. The time complexity of the representation is $O(n^2)$ and for similarity analysis they adopts four element covariance matrix. To represent the graph they initially convert the sequence into a binary string, for example, ATGGTGGA... becomes 001101011101100... where, A, G, C, T as 00, 01, 10, 11, respectively. From the above binary sequence they generate a worm curve [20] (shown in figure 9). After getting the curve, they construct another graph by ignoring the 'zeros' and counting 'ones' as dark spot (See the figure 9 and 10 below)

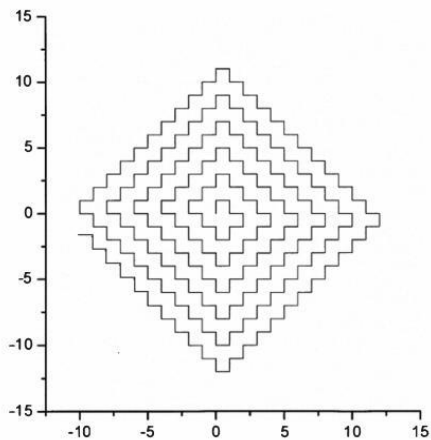

**Fig. 9. Worm curve**

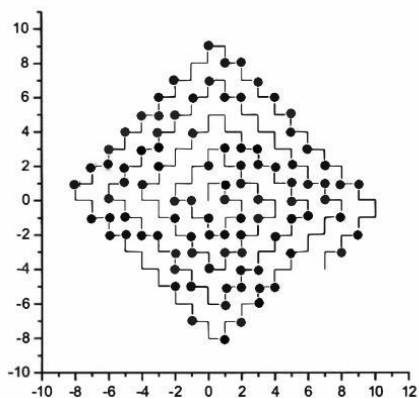

**Fig. 10. The compact graphical representation of β - globin gene (consider only first exon)**

For similarity analysis, dark spots are considered as points and using those points they find four element covariance matrices. Particulars are given as follows –

$$(\mu_a, \mu_b) = (E(A), E(B))$$
$$= \left(\frac{1}{n}\sum_{i=1}^{n} A_i, \frac{1}{n}\sum_{i=1}^{n} B_i\right) \quad (1)$$

$$M_1 = E[(A - \mu_a)(A - \mu_a)]$$
$$= \frac{1}{n}\sum_{i=1}^{n}(A_i - \mu_a)(A_i - \mu_a) \quad (2)$$

$$M_2 = E[(A - \mu_a)(B - \mu_b)]$$
$$= \frac{1}{n}\sum_{i=1}^{n}(A_i - \mu_a)(B_i - \mu_b) \quad (3)$$

$$M_3 = E[(B - \mu_b)(A - \mu_a)]$$
$$= \frac{1}{n}\sum_{i=1}^{n}(B_i - \mu_b)(A_i - \mu_a) \quad (4)$$

$$M_4 = E[(B - \mu_b)(B - \mu_b)]$$
$$= \frac{1}{n}\sum_{i=1}^{n}(B_i - \mu_b)(B_i - \mu_b) \quad (5)$$

Now from the above four element covariance matrix they get a vector as descriptor shown below.

$$\vec{D} = [M_1 M_2 M_3 M_4]^T \quad (6)$$

Then find the Euclidian distance between two vectors $\vec{D}_i$ to $\vec{D}_j$ find similarity between two species i and j respectively.

$$d_{ij} = \vec{D}_i - \vec{D}_j \quad (7)$$

Then they come to a conclusion that, if the distance is smaller than the sequence is high similar.

**In 2011, Xingqin Qi** et al. construct a weighted directed graph for a DNA sequence [21-23]. To construct the graph sequence of nucleotides (A, C, G, T) S is divided into $S_1, S_2, S_3,...,S_n$ where n is the length of the given sequence. Between every pair of vertices they set a weighted edge. Weight of an edge between $S_i$ and $S_j$ is $(1/(j-i)^\alpha)$, where $\alpha > 0$ and $i < j$, $\alpha$ is a parameter which is defined by user to find relationship between smaller distance and bigger distance between two nucleotides. Using this function they construct a weighted directed multi graph $G_m$ (see figure 11) for a sequence S = ACGTATC with $\alpha = 1/2$. There may be some parallel edges in $G_m$ with different weights, replace them with a single edge and sum of weights is assign to the edge. After the simplification a new graph $G_s$ is constructed shown in figure 12.

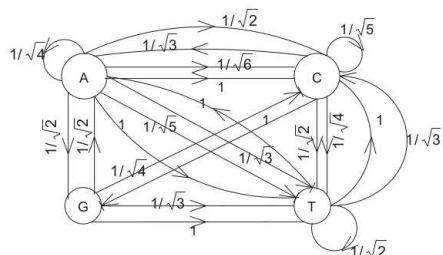

**Fig. 11. Directed multi graph $G_m$ for S = ACGTATC with α = 1/2.**

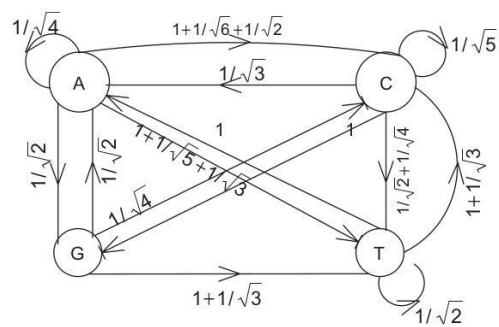

**Fig. 12. $G_s$ for S = ACGTATC after simplification.**

This weighted directed now $G_s$ represent to a 4x4 adjacency matrix M, shown below:

$$M = \begin{pmatrix} (A,A)_w & (A,C)_w & (A,G)_w & (A,T)_w \\ (C,A)_w & (C,C)_w & (C,G)_w & (C,T)_w \\ (G,A)_w & (G,C)_w & (G,G)_w & (G,T)_w \\ (T,A)_w & (T,C)_w & (T,G)_w & (T,T)_w \end{pmatrix}$$

Then M, is also represented in 16 – dimensional vector $\vec{R}^T$ by row major,
$\vec{R}^T = [(A,A)_w, \ldots, (A,T)_w, (C,A)_w, \ldots, (C,T)_w, \ldots, (T,T)_w]$

For the given sequence S = ACGTATC, they got (4x4) matrix and 16 – dimensional matrix also shown as follows:

$$M = \begin{pmatrix} 0.5000 & 2.1154 & 0.7071 & 2.0246 \\ 0.5774 & 0.4472 & 1.0000 & 1.2071 \\ 0.7071 & 0.5000 & 0 & 1.5774 \\ 1.0000 & 0.5000 & 0 & 0.7071 \end{pmatrix}$$

$\vec{R}_s = [0.5000, 2.1154, 0.7071, 2.0246, 0.5774, 0.4472,$
$\qquad 1.0000, 1.2071, \; 0.7071, 0.5000, 0, 1.5774,$
$\qquad 1.0000, 0.5000, 0, 0.7071\;]$

For similarity analysis they used two DNA sequences s and h and represent graphs and get two dimensional vector $\vec{R}_s$ and $\vec{R}_h$ respectively. From these vectors they measure distance $d_1$ by Euclidean Distance measure technique, $d_2$ by using $(1 - \cos)$ of the included angle between $\vec{R}_s$ and $\vec{R}_h$ and $d_3$ by using Pearson Correlation Coefficient technique. Using the abovementioned distance measurement technique following table [24] are constructed for twelve species. Also show the comparison between them using phylogenetic trees [25] (see figure 13).

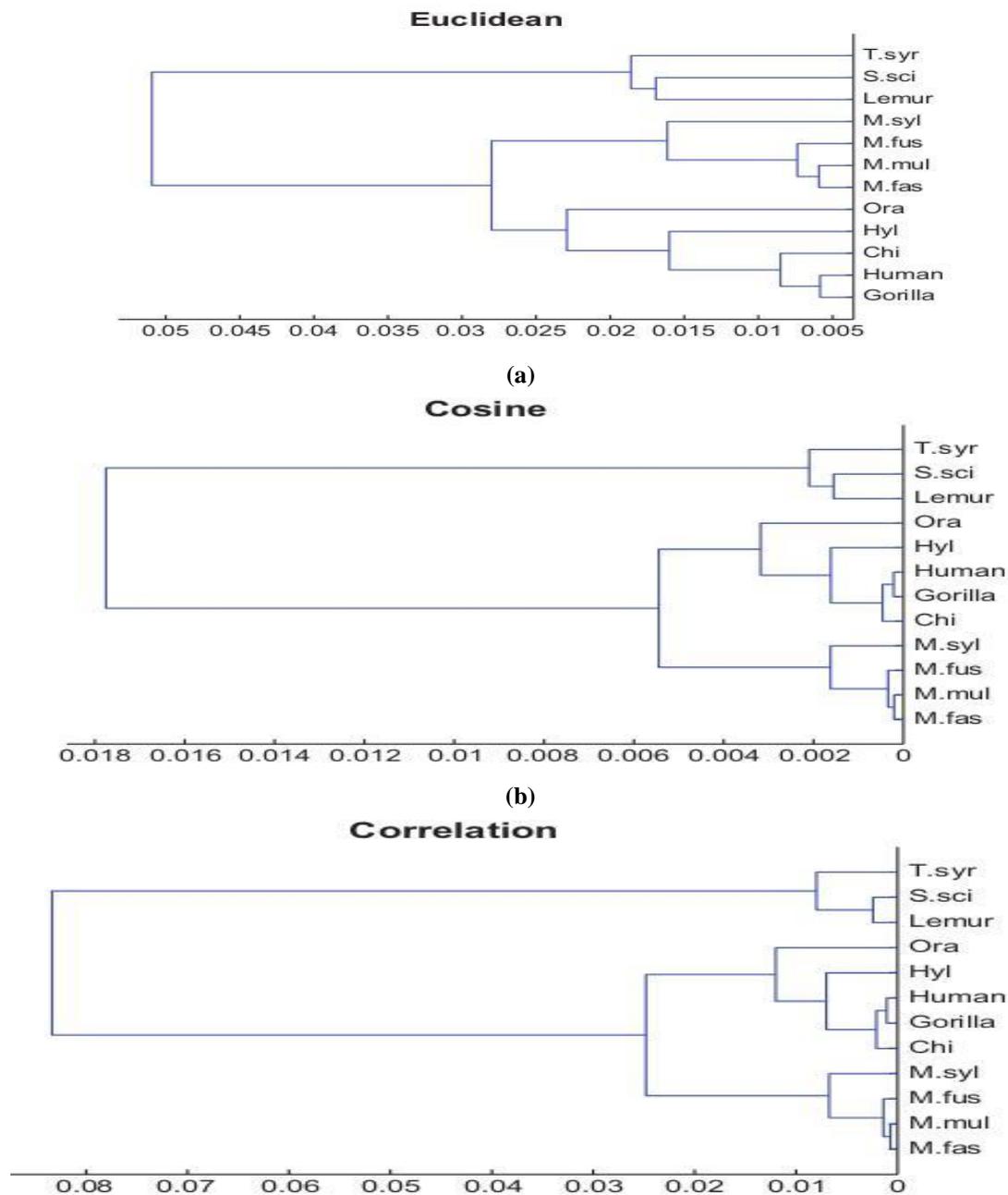

**Fig. 13. Phylogenetic tree on basis of the Tables II –IV.**

**Estimation of Similarity between DNA Sequences and Its Graphical Representation**

Table- II: Similarity/dissimilarity measurement matrix based on $d_1$

| Species | Lemur | Chi | S.sci | M.fas | Gorilla | M.fus | M.mul | M.syl | Hyl | Ora | T.syr | Human |
|---|---|---|---|---|---|---|---|---|---|---|---|---|
| Lemur | 0 | 0.0511 | 0.0169 | 0.0358 | 0.0539 | 0.0373 | 0.0327 | 0.0221 | 0.0510 | 0.0702 | 0.0171 | 0.0591 |
| Chi | | 0 | 0.0528 | 0.0183 | 0.0072 | 0.0171 | 0.0211 | 0.0325 | 0.0179 | 0.0264 | 0.0654 | 0.0098 |
| S.sci | | | 0 | 0.0362 | 0.0545 | 0.0395 | 0.0347 | 0.0286 | 0.0496 | 0.0716 | 0.0201 | 0.0592 |
| M.fas | | | | 0 | 0.0210 | 0.0085 | 0.0059 | 0.0172 | 0.0196 | 0.0391 | 0.0488 | 0.0255 |
| Gorilla | | | | | 0 | 0.0186 | 0.0233 | 0.0354 | 0.0133 | 0.0211 | 0.0679 | 0.0058 |
| M.fus | | | | | | 0 | 0.0063 | 0.0181 | 0.0174 | 0.0342 | 0.0514 | 0.0238 |
| M.mul | | | | | | | 0 | 0.0131 | 0.0212 | 0.0397 | 0.0463 | 0.0284 |
| M.syl | | | | | | | | 0 | 0.0326 | 0.0509 | 0.0355 | 0.0406 |
| Hyl | | | | | | | | | 0 | 0.0243 | 0.0631 | 0.0169 |
| Ora | | | | | | | | | | 0 | 0.0841 | 0.0198 |
| T.syr | | | | | | | | | | | 0 | 0.0730 |
| Human | | | | | | | | | | | | 0 |

Table- III: Similarity/dissimilarity measurement matrix based on $d_2$

| Species | S.sci | Chi | Lemur | M.fas | Gorilla | M.fus | M.mul | M.syl | Hyl | Ora | T.syr | Human |
|---|---|---|---|---|---|---|---|---|---|---|---|---|
| S.sci | 0 | 0.0163 | 0.0016 | 0.0080 | 0.0182 | 0.0087 | 0.0067 | 0.0030 | 0.0163 | 0.0305 | 0.0018 | 0.0219 |
| Chi | | 0 | 0.0177 | 0.0021 | 0.0003 | 0.0018 | 0.0028 | 0.0066 | 0.0020 | 0.0043 | 0.0269 | 0.0006 |
| Lemur | | | 0 | 0.0084 | 0.0190 | 0.0099 | 0.0076 | 0.0050 | 0.0160 | 0.0321 | 0.0024 | 0.0225 |
| M.fas | | | | 0 | 0.0028 | 0.0004 | 0.0002 | 0.0018 | 0.0025 | 0.0094 | 0.0150 | 0.0042 |
| Gorilla | | | | | 0 | 0.0022 | 0.0034 | 0.0078 | 0.0011 | 0.0027 | 0.0290 | 0.0002 |
| M.fus | | | | | | 0 | 0.0002 | 0.0020 | 0.0019 | 0.0072 | 0.0166 | 0.0036 |
| M.mul | | | | | | | 0 | 0.0011 | 0.0028 | 0.0098 | 0.0135 | 0.0051 |
| M.syl | | | | | | | | 0 | 0.0066 | 0.0160 | 0.0079 | 0.0103 |
| Hyl | | | | | | | | | 0 | 0.0034 | 0.0252 | 0.0018 |
| Ora | | | | | | | | | | 0 | 0.0438 | 0.0023 |
| T.syr | | | | | | | | | | | 0 | 0.0336 |
| Human | | | | | | | | | | | | 0 |

Table- IV: Similarity/dissimilarity measurement matrix based on $d_3$

| Species | S.sci | Chi | Lemur | M.fas | Gorilla | M.fus | M.mul | M.syl | Hyl | Ora | T.syr | Human |
|---|---|---|---|---|---|---|---|---|---|---|---|---|
| S.sci | 0 | 0.0749 | 0.0024 | 0.0360 | 0.0840 | 0.0396 | 0.0302 | 0.0136 | 0.752 | 0.1348 | 0.0082 | 0.1015 |
| Chi | | 0 | 0.0869 | 0.0098 | 0.0014 | 0.0087 | 0.0134 | 0.0302 | 0.0081 | 0.0183 | 0.1252 | 0.0027 |
| Lemur | | | 0 | 0.0429 | 0.0959 | 0.0473 | 0.0366 | 0.0196 | 0.0851 | 0.1485 | 0.0078 | 0.1142 |
| M.fas | | | | 0 | 0.0137 | 0.0016 | 0.0007 | 0.0068 | 0.0123 | 0.0409 | 0.0709 | 0.0205 |
| Gorilla | | | | | 0 | 0.0103 | 0.0166 | 0.0358 | 0.0046 | 0.0103 | 0.1367 | 0.0011 |
| M.fus | | | | | | 0 | 0.0012 | 0.0090 | 0.0076 | 0.0316 | 0.0773 | 0.0171 |
| M.mul | | | | | | | 0 | 0.0044 | 0.0127 | 0.0416 | 0.0629 | 0.0247 |
| M.syl | | | | | | | | 0 | 0.0284 | 0.0431 | 0.0357 | 0.0476 |
| Hyl | | | | | | | | | 0 | 0.0708 | 0.1210 | 0.0083 |
| Ora | | | | | | | | | | 0.0109 | 0.1956 | 0.0086 |
| T.syr | | | | | | | | | | | 0 | 0.1586 |
| Human | | | | | | | | | | | | 0 |

## IV. DEMOGRAPHIC ANALYSIS OF GENE EXPRESSION

In the field of bioinformatics, different methods are used for demographical analysis on a gene expression [26]. It is very much needed to know the feature of gene. In relation with environment, important information of gene expression data [27] is needed for understanding biological process. To understand the functional genomics; we need to decipher the hidden patterns of the gene expression data. Reduce the complexity of demographical analysis is difficult due to rapid increase of gene volume and biological network. The clustering of gene expression [28] data is useful to understanding gene functions. It is a useful technique for the identification of similarity, which is very important for design a vaccine. Graph theory also a tool for clustering gene expression data [29] now a days.

## V. STRUCTURAL ANALYSIS OF DNA REPLICATION

DNA replication is also a well known field of biology. Structural analysis of DNA replication [30, 31, 32] explain how it allow to store genetic information within a cell and passing genetic information from one generation of cells to the next. Experiments are currently underway to identify the origins of replication and finding the mutation in the DNA structure. The overall objectives in molecular biology are to identify the key barriers and priorities in the abovementioned list of tasks.

## VI. CONCLUSION

Both DNA sequence and bioinformatics are fast expanding field related research. It is important to identify the area of research in bioinformatics and develop new sequence alignment methods for DNA sequence analysis. Here we have provided a short overview of bioinformatics from a biological sequence analysis perspective. Although a comprehensive survey of all kinds of sequence analysis methods and their effectiveness in bioinformatics is well beyond the task of this short survey. In bioinformatics, Sequence analysis of DNA nucleotides using graph theory has been increased rapidly. It is believed that active interactions and collaborations between these two fields have just started. We hope this study will help accelerate the investigations regarding sequence analysis.

## REFERENCES

1. Luscombe, Nicholas M., Dov Greenbaum, and Gerstein M. "What is bioinformatics? An introduction and overview." Yearbook of Medical Informatics 1.83-100 (2001)
2. Mount, D. W., "Bioinformatics: Sequence and genome analysis". Spring Harbor Press, (2002).
3. Andrew Travers and Georgi Muskhelishvili, "DNA structure and function", FEBS Journal 282, 2279–2295, 2015
4. A. Roy, C. Raychaudhury, A. Nandy, "Novel techniques of graphical representation and analysis of DNA sequences—A review", Volume: 23, Pages: 17, DOI: 10.1007/bf02728525, 1998
5. C. Zhang, R. Zhang, and H. Ou, "The Z curve database: a graphic representation of genomic sequences" Bioinformatics, vol. 19, no. 5, pp.593-599, Mar 2003.
6. R. Zhang, C. T. Zhang, Z curve, "An intuitive tool for visualizing and analyzing the DNA sequences", J. Biomol. Struct. Dyn. 11 (1994) 767–782.
7. E. Hamori, J. Ruskin, H curves, "A novel method of representation of nucleotide series specially suited for long DNA sequences", J. Biol. Chem. 258, 1318–1327, 1983
8. Sai Zou, Lei Wang, Junfeng Wang, "A 2D graphical representation of the sequences of DNA based on triplets and its application", EURASIP Journal Bioinformatics Systems Biology, DOI:10.1186/16874153, 2014
9. Yue Huang and Ling Zhang, "Rapid and sensitive dot-matrix methods for genome analysis", Bioinformatics, Vol. 20 no. 4 2004, pages 460–466, 2003
10. Adrian J. Gibbs, George A. Mcintyre, "The Diagram, a Method for Comparing Sequences : Its Use with Amino Acid and Nucleotide Sequences ", Volume: 16, DOI: 10.1111/j, pg. 1432-1033, 1970.
11. Needleman, S. B., and Wunsch, C. D, "A general method applicable to the search for similarities in the amino acid sequence of two proteins". J Mol Bio l 48, 443–53, 1970.
12. T. F. Smith and M. S. Waterman, "Identification of Common Molecular Subsequences." Journal of Molecular Biology, vol. 147, no. 1, pp. 195–197, 1981. [Online].
13. Altschul SF, Gish W, Miller W, Myers EW, Lipman DJ. "Basic Local Alignment Search Tool", J Mol Biol, 215:403–410, 1990
14. Kit J. Menlove, Mark Clement, and Keith A. Crandall, "Similarity Searching Using BLAST" Journal of Bioinformatics for DNA Sequence Analysis, Methods in Molecular Biology 537, DOI 10.1007/978−1−59745−251−91, 2009
15. Altschul SF, Madden TL, Schaffer AA, Zhang J, Zhang Z, Miller W, Lipman DJ. Gapped BLAST and PSI-BLAST: "A new generation of protein database search programs", Nucleic Acids Res, 25:3389– 3402, 1997
16. Pearson WR, Lipman DJ. "Improved tools for biological sequence comparison", Proc Natl Acad Sci USA 85: 2444-2448, 1988
17. Liu, K., Warnow, T. J., Holder, M. T., Nelesen, S. M., Yu, J., Stamatakis, A. P., Linder, C. R., " SAT´e-II: "Very Fast and Accurate Simultaneous Estimation of Multiple Sequence Alignments and Phylogenetic Trees", Journal of Society of Systematic Biologists, DOI : 10.1093/sysbio/syr095, June 2011
18. Jia-Feng Yu, Ji-Hua Wang, Xiao Sun, "Analysis of Similarities/Dissimilarities of DNA Sequence Based on a Novel graphical Representation", MATCH Communications in Mathematical and in Computer chemistry, 2010
19. Zhujin Zhang, Shuo Wang, Xingyi Zhang,Zheng Zhang, "Similarity analysis of DNA sequences based on a compact representation", 978-1-4244-6439, 2010 IEEE
20. M. Randic, M. Vranco, J. Zupan, and M. Novic, "Compact 2-D graphical representation of DNA", Chem. Phys. Lett. vol. 373, no. 5-6, pp. 558-562, May 28 2003.
21. Xingqin Qi, Qin Wu, Yusen Zhang, Eddie Fuller and Cun-Quan Zhang, "A Novel Model for DNA Sequence Similarity Analysis Based on Graph Theory", Evolutionary Bioinformatics 2011:7 149–158.
22. Anjaneyulu G.S.G.N , A. Kamboj, S. Kalra, "DNA Sequencing Similarity Analysis: Graph theory Application", IJSER, Volume 6, October 2015
23. N. Jafarzadeh, A. Iranmanesh, "A New Graph Theoretical Method for Analyzing DNA Sequences Based on Genetic Codes", MATCH Commun. Math. Comput. Chem. 75(2016) 731-742, 2015
24. B. Liao, W. Zhu, Y. Liu, "3D graphical representation of DNA sequence without degeneracy and its applications in constructing phylogenic tree", MATCH Commun. Math. Comput. Chem. 56, 209–216, 2006.
25. A. Nandy, "A New Graphical Representation and Analysis of DNA Sequence Structure: I. Methodology and Application to Globin Genes", Curr. Sc. 66, 309-314, 1994
26. Oyelade, J., Isewon, I., Oladipupo, F., Aromolaran, O., Uwoghiren, E., Ameh, F., Achas, M. and Adebiyi, E., "Clustering algorithms: Their application to gene expression data", Bioinformatics and Biology Insights, Vol. 10, pp. 237-253, 2016.
27. Pirim H, Ek¸sio˘glu B, Perkins AD, Yu¨ceer C¸., "Clustering of high throughput gene expression data", Computer Operation Research, Vol. 39, issue 12, pp. 3046–3061, 2012.
28. Ying Xu, Victor Olman, Dong Xu, "Clustering gene expression data using a graph-theoretic approach: an application of minimum spanning trees", Bioinformatics, Volume 18, Issue 4, Pages 536–545, 2002
29. John Herrick, Aaron Bensimon, "Single molecule analysis of DNA replication", Biochimie 81, (859− 871), 1999
30. Chevalier S., Blow J.J., "Cell cycle control of replication initiation in eukaryotes", Curr. Opin. Cell Biol.8, 815–821, 1996
31. Campbell J.L., Yeast "DNA replication", J. Biol. Chem. 268, 25261–25264, 1993.

32. Book: iGenetics A Molecular Approach by Peter J. Russel, 3rd Edition, Pearson

## AUTHORS PROFILE

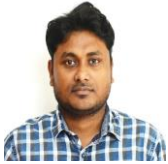

**Probir Mondal,** M. Sc.**,** M. tech in Computer Science and Technology. Presently posted as an Assistant professor in the Department of Computer Science, P. R. Thakur Govt. College, West Bengal, India. Area of interest is Graph Theory, Bioinformatics and Analysis of Algorithm.